\newcommand{\rem}[1]{}
\documentclass[prd,twocolumn,showpacs,showkeys,showpacs,preprintnumbers,nofootinbib,amsmath,amssymb]{revtex4-2}
\usepackage{amsfonts,amssymb,amsmath,mathrsfs}
\usepackage{url}
\usepackage{graphicx}
\newtheorem{thrm}{Theorem}
\usepackage[colorlinks,citecolor=blue,urlcolor=blue,linkcolor=blue]{hyperref}

\newtheorem{remark}[thrm]{Remark}
\usepackage{xcolor}

\begin{document}
\title[Cosmological parameters for CKG]{Constraints on cosmological parameters\\and CMB first acoustic peak in conformal Killing gravity}
\author{Salvatore Capozziello$^{1,2}$}
\email{capozziello@na.infn.it}
\author{Carlo Alberto Mantica$^3$}
\email{carlo.mantica@mi.infn.it} 
\author{Luca Guido Molinari$^{3}$} 
\email{luca.molinari@mi.infn.it}
\author{Giuseppe Sarracino$^{4}$}
\email{giuseppe.sarracino@inaf.it}
\address{$^1$Dipartimento di Fisica  E. Pancini, Universit\`a degli Studi di Napoli Federico II,
Napoli\\ and Istituto Nazionale di Fisica Nucleare (INFN),  Sez di Napoli, Compl. Univ. di Monte S. Angelo,
Edificio G, Via Cinthia, 80126 Napoli, Italy}
\address{$^2$Scuola Superiore Meridionale, Largo S. Marcellino 10,
 80138 Napoli, Italy}
\address{$^3$Dipartimento di Fisica  Aldo Pontremoli, Universit\`a degli Studi di Milano
and INFN,  Sez, di Milano, Via Celoria 16, 20133 Milano, Italy}
\address{$^4$Osservatorio Astronomico di Capodimonte (OACN), Istituto Nazionale
di Astrofisica (INAF), Salita Moiariello 16, 80131 Napoli, Italy}
\date{\today}

\begin{abstract}
In the frame of conformal Killing gravity cosmology, we performed a Bayesian analysis on two different datasets of Baryon Acoustic oscillations (DESI and SDSS DR16), two datasets of SNeIa (Pantheon+ and Union3), and using the Cosmic Microwave Background (CMB) Planck likelihood. The results for $H_0$ and $\Omega_M$ in a spatially flat Friedmann-Lema\^itre-Robertson-Walker (FLRW)  background are consistent with the $\Lambda$CDM scenario. We obtain a non-negligible negative value for the novel density of dark sector, $\Omega_D$, and its relevance in the evolution of the cosmological observables,
thus finding quantitatively what its contribution is on real data to match the standard scenario. The results
confirm the dynamical character of dark energy. We also calculate
the deceleration parameter $q_0$ and the present time dark energy
equation of state parameter $w_0$: the latter belongs to the quintessence regime.
The evaluation of the first acoustic peak of CMB places it
near to the best value provided by the Planck collaboration. In this scenario, we can conclude that late time and early time data can be successfully matched under the same standard.
\end{abstract}

\date{\today}


\maketitle

\section{Introduction}
In the last decades General Relativity (GR) and deep space surveys have radically changed our understanding and view of the Universe. Since its beginning,  GR has predicted and has been confirmed in crucial experiments: from early ones, such 
as the precession anomaly of Mercury, the bending of light, gravitational redshift, to recent ones, as gravitational lensing \cite{Will 06}, the black hole shadow, gravitational waves from compact mergers \cite{Abbott 16}. \\
Despite the successes, GR encounters challenges that precision measures make compelling:
the flat rotation curves of stars in galaxies and the stability of galactic clusters (the dark matter puzzle), the present
accelerated phase of the Universe (the dark energy puzzle) \cite{Riess 98,Perlmutter 99}. The standard cosmological model faces two mainstreams: either introduce new forms of matter and energy, or modify the gravitational field equations \cite{CDL11,DeFelice}.

In the first instance, the cosmological constant $\Lambda$ 
has been the primary candidate for dark energy \cite{Weinberg 89}, but it meets difficulties, like the {\em fine-tuning problem}\footnote{The fundamental constants match the stability range for stars and planets and for life to form.} and the {\em coincidence problem}\footnote{The dark matter and energy densities are of the same order of magnitude.}   \cite{Pad 03,Copeland 06}. For this reason\textcolor{red}{,} other dark energy models have
been proposed, such as quintessence \cite{Ratra 88, Capozziello}, phantom energy \cite{Caldwell03} oscillating quintom \cite{Feng 06}, ghost \cite{Schutzold 02}, Chaplygin gas \cite{Kamenshchik 01}.

The alternatives  are the theories of modified gravity, where the shortcomings of GR in fitting 
astrophysical and cosmological structures require more ``geometric" degrees of freedom 
that modify the gravitational interaction at large scale and avoid the huge, yet undetected, amount of exotic matter  \cite{Capozziello19}. Some popular models are: 
scalar-tensor theories \cite{Bergmann 68}, Einstein-aether
theory \cite{Jacobson 01}, $f(R)$ gravity \cite{Sotiriu10,Odi1,Capozziello07,Odi2}, Gauss - Bonnet gravity \cite{Bajardi,NojOd05}, mimetic gravity \cite{Chamseddine 13}. For reviews see \cite{CDL11,CF08}. Dark matter and dark energy can be both investigated in $f(R)$ framework as discussed in  \cite{DarkMatt, Odi 3}.

A breakthrough in the comprehension of dark energy are the results of the Dark Energy Spectroscopic Instrument (DESI) \cite{Adame 25 DESI,Adame 25 b,Abdul DESI release 2}.
In particular, Ref.  \cite{Adame 25 DESI} provides the transverse comoving distances and the Hubble rates of over 6 million extragalactic objects in the redshift range $0.1<z<4.2$. The dark energy dynamics is tested
in the flat $w$CDM model assuming an equation of state (EoS) for dark energy $p_D=w \mu_D$ with constant $w$, and in the $w_0 w_a$CDM model, where the EoS is linearly scale-dependent: $w=w_0+w_a(1-a)$. 
The DESI data release 2 presents measures of baryon acoustic oscillation (BAO) 
in more than 14 million galaxies and quasars \cite{Abdul DESI release 2}.

A large literature eventually flourished, with main focus on the persistent deviations of DESI data from standard 
$\Lambda$CDM cosmology (see for example \cite{Li 24,Wen24,Malekjani 25,Barua 25,Chudaykin 24,Berti 25,Giare24a,Giare24b,Yang 25,Gialamas 25,Colgain,Wang25}).
Several papers test the evolution of dark energy 
parametric models for $w(a)$, like the one by Chevallier-Polarski-Linder
\cite{Chevallier01,Linder 03} or by Jassal-Bagla-Padmanabhan 
 \cite{Jassal 05}, with experimental data.

In \cite{Giare24b}, the implications of BAO measurements by DESI are investigated 
for models with energy-momentum flow among dark matter and dark energy. In \cite{Berti 25},
a non-parametric, model independent reconstruction of the dark energy
density evolution, using DESI, is performed. In \cite{Odi 3-1}, a generalisation
of exponential $f(R)$ gravity is considered and compared with CDM
model by using the latest data from DESI. The analysis suggests that the 
model provides much better fits than CDM model. A similar result is achieved in \cite{Himanshu} where data seem to challenge the standard $\Lambda$CDM paradigm.

In this context, Harada \cite{Harada23,Harada23b} introduced the Conformal Killing gravity (CKG) in 2023 
 to face the issue of dark energy. 
It is defined by the following field equations, that are third order in the derivatives of the 
metric tensor\footnote{We use Latin indices for space-time components, and Greek indices for space components.}:
\begin{align}
H_{jkl}=&\,T_{jkl} \label{eq:Harada first}\\
H_{jkl}=&\nabla_{j}R_{kl}+\nabla_{k}R_{lj}+\nabla_{l}R_{jk}\\
&-\tfrac{1}{3}(g_{kl}\nabla_{j}R+g_{lj}\nabla_{k}R+g_{jk}\nabla_{l}R)\nonumber \\
T_{jkl}=&\nabla_{j}T_{kl}+\nabla_{k}T_{lj}+\nabla_{l}T_{jk}\\
&-\tfrac{1}{6}(g_{kl}\nabla_{j}T+g_{lj}\nabla_{k}T+g_{jk}\nabla_{l}T) \nonumber
\end{align}
$R_{jk}$ is the Ricci tensor with trace $R$, $T_{kl}$ is
the stress-energy tensor with trace $T$.\\ 
The Bianchi identity $\nabla_{j}R^{j}{}_{k}=\frac{1}{2}\nabla_{k}R$
implies $\nabla_{j}T^{j}{}_{k}=0$. Solutions of the Einstein equations
are solutions of the new theory. The cosmological constant
$\Lambda$ arises as an integration constant. 

Soon after, Mantica and Molinari \cite{Mantica 23 a-1}
found  a parametrization showing that the Harada equations are equivalent to
the Einstein equations modified by a supplemental conformal Killing
tensor that is divergence-free: 
\begin{align}
 & R_{kl}-\tfrac{1}{2}Rg_{kl}= \, T_{kl}+K_{kl} \label{eq:einstein enlarged}\\
 & \nabla_{j}K_{kl}+\nabla_{k}K_{jl}+\nabla_{l}K_{jk} \nonumber\\
 &=\tfrac{1}{6}(g_{kl}\nabla_{j}K+g_{jl}\nabla_{k}K+g_{jk}\nabla_{l}K)\label{eq:Conformal Killing mantica-1}
\end{align}
The name of the theory is after this feature.
The reformulation makes the extension of GR explicit through the
conformal Killing term, that satisfies $\nabla^{k}K_{kl}=0$. It appears in the equation as a source term, 
and suggests itself as a candidate for the dark sector.\\
Geometric properties of conformal Killing tensors are discussed in refs.\cite{geom}.

Most of the studies in CKG are in static spacetimes. Let us recap some important ones: Barnes \cite{Barnes23a}
found the most general spherically symmetric solution both in vacuum
and with a Maxwell source \cite{Barnes 23b}. Junior, Lobo and Rodriguez
 investigated regular black hole solutions \cite{Junior24,Ednaldo J 25}, as well as black bounce  
 coupled to non-linear electrodynamics and scalar fields \cite{Junior 24 b}. 
Dynamics of thin-shell wormholes made of two CKG
black holes was pursued in \cite{Alshal 24}. 

In Ref. \cite{Mantica 24b}, Mantica and Molinari gave a parametrization of CKG
in static spherically symmetric background, with a conformal Killing tensor of the form of energy-momentum for 
an anisotropic fluid. The fact that the field equations turn to second order considerably simplifies 
the search of solutions with respect to the original formulation.
The same authors derived a Tolman-Oppenheimer-Volkoff
equation for CKG in static spherically symmetric spacetimes, and a solution
for a perfect fluid sphere with uniform matter density \cite{Mantica 25 }.

In \cite{Altas 25} Altas and Tekin claimed that 
field equations based on a tensor with rank greater than 2 are problematic
for the interpretation of integration constants,
such as the parameter $m$ in Schwarzchild's metric, in terms of conserved quantities of the theory.
Starting from Harada's equations they showed that $\nabla^{k}\nabla^{j}H_{jkl}=\nabla^{k}\Phi_{kl}$, 
and argued that the current  $J_{k}=\Phi_{kl}\xi^{l}$ inferred via a Killing vector
is zero for Einstein spaces and for the Schwarzschild metric. In \cite{Mantica 25 b}
the authors verified that  $J_k$ is actually nonzero for Harada's vacuum
solution and, in static spherically
symmetry as well as in cosmological background, 
CKG is (because of parametrization)
not third order  \cite{Mantica 24b,Mantica 25 }. This permits to introduce
conserved currents in CKG 
as standard contractions of the divergence-free stress-energy and of the conformal Killing tensors
with the available Killing vectors.

It is worth mentioning  the 
works on wave and Kundt metrics in CKG \cite{GGurses 24,Hervik 24},
and the discussion of the cosmological constant \cite{Feng 24}.

Only a few papers have been devoted so far to cosmology in
CKG. Vacuum cosmological solutions
were discussed by Clem\'ent and Nouicer \cite{Clement24} together with
wormhole and black hole ones. 
In Ref. \cite{Mantica 23 a-1} a conformal Killing parametrization was obtained in a Robertson-Walker (RW) background: a perfect fluid $K_{jl}$ describing the dark sector.
The Friedmann equations with the Killing source gave the same forecast
for the dark fluid as Harada in \cite{Harada23b}.
In particular,  a $\Lambda$ term arises, and the field
equations are second order.  

The first attempt to obtain quantitative forecasts of cosmological
parameters in CKG was developed in \cite{Mantica 24}. The analytic expression of the Hubble function 
versus redshift $H(z)$ was fitted 
with cosmological data based on cosmic chronometers (CC) or CC and
baryon acoustic oscillations (CC+BAO). The dark energy density and pressure were found to
vary with the scale, with large dependence on the sign of $\Omega_D$.\\
Similar dichotomic results were obtained for the function $\sigma_{8}(z)$: it was shown that the negative value of $\Omega_D$ implies a peak of the function $g(z)$, sharper than the peak in $\Lambda$CDM, both in the range $0<z<1$, while $\Omega_D>0$ gives a local plateau in $z=0$. 
\begin{table}[h]
\centering 
\begin{tabular}{|c|c|c||c|c|c|c|}
\hline
& $\Lambda$CDM & CKG &  $\Lambda$CDM & CKG\\
\hline
$H_0$                       & 68.16 & 71.42     & 69.83 &64.17\\
$q_0$ & -0.516 &-0.854 &-0.549 & -0.130\\ 
$\Omega_M$             & 0.323 & 0.311  & 0.301 & \,0.306 \\
$\Omega_\Lambda$  & 0.677 & 0.368   &0.699 & \,1.103\\
$\Omega_D$             & --        & +0.321  & --         & -0.410\\
Age                            &13.59  & 13.47   & 13.92 &\,14.10 \\
\hline
\end{tabular}
\caption{The central values of the cosmological parameters for $\Lambda$CDM and CKG evaluated in 
Ref.\cite{Mantica 24} by fit with CC (left) and CC+BAO (right) data. Note the opposite signs of $\Omega_D$ and the relative stability of $\Omega_M$.}
\label{Oldtable}
\end{table}
The equations in the linear approximation for the evolution of the density contrast in a matter dominated Universe were also solved: the dark
and the $\Lambda$ terms gave no significant deviation from the $\Lambda$CDM
results. The role of dark energy remained quite elusive.

In the present paper,  we clarify this ambiguity. We use the BAO datasets
from DESI DR2 and the Sloan Digital Sky Survey (SDSS) DR16 combined with the Supernovae type Ia (SNeIa) data of Pantheon+ and Union3 to fix the cosmological parameters of CKG in a spatially flat RW background. We complement these results by also including the Cosmic Microwave Background (CMB) Planck \cite{Planck 2018} likelihood, and drop the assumption of flatness for one specific computation.

The paper is organized as follows. 
In Section \ref{sec:Friedmann-equations} we shortly recap the Friedmann
equations as they appear in CKG. In Section \ref{sec:data_analysis} we present the datasets utilised in this work. Section \ref{sec:results} is devoted to the methodology and the main cosmological results for the CKG parameters.
In Section \ref{perturbations} some considerations are drawn for the growth of perturbations in CKG cosmology and of $\sigma_8$. Discussions and conclusions are drawn in Section \ref{sec:discussion and conclusions}.

\section{The Friedmann equations\\ in conformal Killing gravity}\label{sec:Friedmann-equations}
The cosmological principle fixes the space-time metric as Robertson-Walker (RW)
\begin{align}
ds^2 = - dt^2 + a^2(t) \left [ \frac{dr^2}{1-kr^2}+ r^2 ( d \theta^2 +\sin^2\theta d \phi^2) \right ] \label{eq:1}
\end{align}
where $a(t)$ is the scale factor and $k=0,\pm1 $ is the parameter for the space curvature. 
An alternative covariant characterization is through the existence of a unit time-like
vector field, $u_{k}u^{k}=-1$, such that (see \cite{Capozziello 22}):
\begin{align}
 & \nabla_{j}u_{k}=H(g_{jk}+u_{j}u_{k}),\label{eq:torse-forming}\\
 & \nabla_j H = -\dot H u_j \label{DOTH}
\end{align}
with zero Weyl tensor.\\ 
In the coordinates \eqref{eq:1},  $H=\dot a/a$ is the Hubble parameter, $u^0=-1$ and $u^\mu=0$. The dot means  
$u^k\nabla_k$, i.e. a derivative in the cosmic time $t$ (for a scalar field $\dot f = u^k\nabla_k f =\partial_t f$).\\
Eq.\eqref{DOTH} is equivalent to $R_{ij}u^j =\xi u_i$, with eigenvalue proportional to the cosmic acceleration
 \begin{align}
 \xi=3(H^2+\dot H )= 3 \frac{\ddot a}{a}\label{xiACC}
 \end{align}
In a RW spacetime the Ricci tensor and the scalar curvature are 
\begin{align}
 & R_{kl}=\frac{R-4\xi}{3}u_k u_l +\frac{R-\xi}{3}g_{kl}\label{eq:2.3 Ricci GRW}\\
 & R=\frac{R^{\star}}{a^{2}}+6H^{2}+2\xi\label{eq:2.9 scalar R}
\end{align}
$R^{\star}=6k$ is the curvature of the spacelike hypersurface. In a RW spacetime, the divergence-free conformal
Killing tensor is found to be 
\begin{align}
 K_{kl}=g_{kl}\left[\frac{5}{6}Ca^{2}-\Lambda\right]+u_{k}u_{l}\frac{Ca^{2}}{3}\label{eq:Conformal Killing GRW}
\end{align}
being $C$ and $\Lambda$ two integration constants \cite{Mantica 23 a-1,Mantica 24}. 

\begin{remark}
Eq.(\ref{eq:Conformal Killing GRW}) is inherited from the conformal Killing symmetry of a RW spacetime. It is a matter of fact that the spacetime is endowed by a unique timelike conformal Killing vector  
$\xi_{j}=a(t) u_{j}$, i.e.
$$\nabla_{i}\xi_{j}+\nabla_{j}\xi_{i}=2\dot a g_{ij}$$
The associated tensor $K_{ij}=\frac{1}{3}C \xi_{i}\xi_{j}+\beta(t) g_{ij}$ 
with arbitrary constant $C$ and function $\beta (t)$ is, by construction,
a conformal Killing tensor. The divergence-free condition imposes $\dot\beta = (5/3)a\dot a$. Integration gives another arbitrary constant: $\beta(t) = \frac{5}{6}C a^2(t)-\Lambda $. Eq.(\ref{eq:Conformal Killing GRW})
is recovered.
\end{remark}

The perfect-fluid tensor (\ref{eq:Conformal Killing GRW}) is
purely geometric and supposedly describes the {\em dark
perfect fluid}: $K_{kl}=(p_{D}+\mu_{D})u_{i}u_{j}+p_{D}g_{ij}$, with 
dark energy density and dark pressure 
\begin{align}
 & \mu_{D}=-\frac{1}{2}Ca^{2}+\Lambda\label{MUD}\\
 & p_{D}=+\frac{5}{6}Ca^{2}-\Lambda\label{PD}
\end{align}
Remarkably,  in CKG,  they depend on the scale (i.e. cosmic time). 
The EoS parameter is:
\begin{equation}
w_{D}(a)=\dfrac{p_{D}}{\mu_{D}}=-1+\dfrac{2Ca^{2}}{6\Lambda-3Ca^2}\label{eq:w(a)}
\end{equation}
While the paramerizations of $w(a)$ by Chevallier-Polarski-Linder 
\cite{Chevallier01,Linder 03} or Jassal-Bagla-Padmanabhan \cite{Jassal 05}
or others in literature, are dictated by convenience to mimic observed data,
the expression \eqref{eq:w(a)} is a direct consequence of   CKG. 

Eqs. (\ref{eq:2.3 Ricci GRW}) and (\ref{eq:Conformal Killing GRW})
give the perfect fluid energy-momentum tensor of ordinary matter
with energy density $\mu$ and pressure $p$: 
\begin{align}
T_{kl}  & =R_{kl}-\tfrac{1}{2}Rg_{kl}-K_{kl}\nonumber \\
 & =-\tfrac{1}{6}(R+2\xi+5Ca^{2}-6\Lambda)g_{kl}\nonumber\\
 & \quad +\tfrac{1}{3}\left(R-4\xi-Ca^{2}\right)u_{k}u_{l}\nonumber\\
& \equiv  (p+\mu)u_{k}u_{l}+pg_{kl} \label{eq:stress energy conformal killing}
\end{align}
After specifying $R$ with (\ref{eq:2.9 scalar R}) and $\xi$ with \eqref{xiACC},
the Friedmann equations for CKG with ordinary matter are: 
\begin{align}
 & \mu=\frac{R^{\star}}{2a^{2}}+3H^{2}+\frac{1}{2}Ca^{2}-\Lambda\label{KAPPAMU}\\
 & p=-\frac{R^{\star}}{6a^{2}}-3H^{2}-2\dot{H}-\frac{5}{6}Ca^{2}+\Lambda\label{eq:Friedmann conformal killing}
\end{align}
With $C=0$ they reduce to the standard GR equations with cosmological
constant $\Lambda$. 

In the standard analysis, ordinary matter is composed of pressure-less dust with energy density $\mu_{M}$,
and radiation with $p_R=\frac{1}{3}\mu_R$. Conservation in the expanding Universe gives:
$\mu_M /a^3 =\mu_{M0} /a_0^3$ and $\mu_R/a^4 = \mu_{R0}/a_0^4$.
The Friedmann Eq. (\ref{KAPPAMU}) of CKG with energy density $\mu=\mu_M+\mu_R$, 
the dark energy (with $\Lambda$ term) and the curvature term is \cite{Mantica 24}
\begin{equation}
\mu_{M0}\left(\frac{a}{a_{0}}\right)^{-3}+\mu_{R0}\left(\frac{a}{a_{0}}\right)^{-4}=\frac{R^{\star}}{2a^{2}}+3H^{2}+\frac{1}{2}Ca^{2}-\Lambda
\end{equation}
Divide by $3H_{0}^{2}$ (the value at scale $a_{0}$) and obtain:
\begin{align}
\left(\frac{H}{H_0}\right)^2 =& \Omega_R \left (\frac{a}{a_0}\right)^{-4} + \Omega_M \left(\frac{a}{a_0}\right )^{-3} \nonumber\\
&+\Omega_k \left(\frac{a}{a_0}\right)^{-2}+\Omega_{\Lambda}+\Omega_{D}\left(\frac{a}{a_0}\right)^{2}\label{eq:ALL H}
\end{align}
\begin{gather*}
\Omega_M = \frac{\mu_{M0}}{3H_0^2}, \quad \Omega_R=\frac{\mu_{R0}}{3H_{0}^{2}},\quad \Omega_{k}=-\frac{R^{\star}}{6H_{0}^{2}a_{0}^{2}},\\
\Omega_{\Lambda}=\frac{\Lambda}{3H_{0}^{2}},\quad \Omega_{D}=-\frac{Ca_{0}^{2}}{6H_{0}^{2}}\\
\Omega_M+\Omega_R+\Omega_{k}+\Omega_{\Lambda}+\Omega_{D}=1
\end{gather*}
While $\Omega_M$ and $\Omega_R$ are true
energy densities and are positive, $\Omega_{\Lambda}$, $\Omega_k$
and $\Omega_{D}$ have geometric origin and their sign is not automatically
positive.

Eq.\eqref{eq:ALL H}, with  $H=\dot a/a$,
gives the time evolution of the scale function $a(t)$. In terms of the redshift parameter $1+z=a_{0}/a$, it becomes: 
\begin{align}
\left(\frac{H(z)}{H_{0}}\right)^{2}=&\Omega_R(1+z)^{4}+\Omega_M(1+z)^{3} \nonumber\\
&+\Omega_{k}(1+z)^{2}+\Omega_{\Lambda}+\frac{\Omega_{D}}{(1+z)^{2}}\label{HSQ}
\end{align}

\subsection{Spatially flat FLRW}
We consider a spatially flat FLRW space-time ($\Omega_k=0$), and neglect the vanishing contribution of radiation in the late Universe. The relevant terms in the equation for $H(z)$ are
\begin{equation}
\left(\frac{H(z)}{H_{0}}\right)^{2}=\Omega_M(1+z)^{3}+\Omega_{\Lambda}+\frac{\Omega_{D}}{(1+z)^{2}}\label{eq:Hubble late time}
\end{equation}
Besides $H$ one may consider the cosmographic parameters ``deceleration''
$q$, ``jerk'' $j$ and ``snap'' $s$ (see \cite{Visser 04}) through the expansion
\begin{align}
\frac{a(t)}{a_0} =& 1 + H_0 (t-t_0) - \frac{H_0^2}{2} q_0 (t-t_0)^2 \nonumber \\
&+\frac{H_0^3}{3!} j_0 (t-t_0)^3 + \frac{H_0^4}{4!} s_0 (t-t_0)^4+\dots
\end{align}
The deceleration is $q=- a\ddot{a}/\dot{a}^2 =-1- (\dot H/H^2)$. With 
$\dot{H}=(dH/dz)\dot{z}$  and $\dot{z}=-H(z+1)$, it is 
\begin{align}
q(z) 
=\frac{\Omega_M(1+z)^{5}-2\Omega_{\Lambda}(1+z)^{2}-4\Omega_{D}}{2\Omega_M(1+z)^{5}+2\Omega_{\Lambda}(1+z)^{2}+2\Omega_{D}} \label{QZ}
\end{align}

\begin{remark}
The contraction $R_{ij} u^i u^j = -\xi = -3 \ddot a/a = 3 H^2 q $ is related to the
``strong energy condition", which is the physical requirement
$$(T_{ij}-\tfrac{1}{2} T g_{ij} )u^i u^j \ge 0$$ 
For a perfect fluid it is $3p+\mu\ge 0$. \\
If $R_{ij} =T_{ij}-\frac{1}{2}Tg_{ij} $, the condition $3H^2 q \ge 0$ contradicts the present time 
acceleration. This is overturned in standard cosmology by the introduction of the cosmological constant, with negative pressure.\\ 
In CKG cosmology the source is $T_{ij}+K_{ij}$. Considering radiation, dust, the dark fluid, and $R^\star=0$, the parameter $q$ is eq.\eqref{QZ}. In Table \ref{Tab_Results_2} we show that $q_0$ is negative for all datasets.\\
The ``weak energy condition" $T_{ij}u^iu^j\ge 0$ means that the energy density $T^{00}$ is positive 
in the comoving frame.
Neglecting radiation it is $\mu_M+\mu_D\ge 0$ i.e. 
$$\Omega_M (1+z)^3+\Omega_\Lambda +\Omega_D (1+z)^{-2} \ge 0 $$
By eq.\eqref{eq:Hubble late time} this is the requirement $H(z)^2\ge 0$. The appearance of a zero $H(z)=0$ is discussed in the next remark.
\end{remark}

The ``lookback time'' of a photon emitted at time $t_{e}$ at 
scale $a_e$, and received at time $t_{0}$ at
 scale $a_{0}$, is: 
\begin{equation}
t_{L}=\int_{t_{e}}^{t_{0}}dt=\int_{a_{e}}^{a_{0}}\frac{da}{\dot{a}}=\int_{0}^{z_{e}}\frac{dz}{H(z)(1+z)} \label{LKBT}
\end{equation}
The age of the Universe is the limit (whenever it exists)
$z\rightarrow+\infty$ of the previous expression. 

The dark energy density (\ref{MUD}) and the dark pressure (\ref{PD})
versus redshift are 
\begin{equation}
\begin{array}{lc}
\mu_{D}= 3H_{0}^{2}\left[\Omega_{\Lambda}+\dfrac{\Omega_{D}}{(1+z)^{2}}\right]\\
\\
p_{D}=-3H_{0}^{2}\left[\Omega_{\Lambda}+\dfrac{5\Omega_{D}}{3(1+z)^{2}}\right]
\end{array}\label{eq:dark pressure and energy}
\end{equation}
with EoS parameter  (\ref{eq:w(a)})
\begin{equation}
w_D(z)=
-1-\dfrac{2}{3}\dfrac{\Omega_{D}}{\Omega_{D}+\Omega_{\Lambda}(1+z)^{2}}\label{eq:EoS in CKG}
\end{equation}
The first order expansion in the scale parameter determines the empirical coefficients of the Chevallier-Polarski-Linder model \citep{Chevallier01,Linder 03}:
\begin{align}
w_D (a) &= w_{D0} + w_{Da} (1-a) \nonumber \\ 
=& -\frac{1}{3}\frac{5\Omega_D+{\bf 3}\Omega_\Lambda}{\Omega_\Lambda + \Omega_D}  + \frac{4}{3}
\frac{\Omega_D\Omega_\Lambda}{(\Omega_D+\Omega_\Lambda)^2} (1-a) \label{JBPinCKG}\,.
\end{align}
See Table \ref{Tab_Results_2} for the numerical values.

\begin{remark}
The equations show a limit value $z=-1$, at which $p_D$ and $\mu_D$ diverge with ratio 
$w_D(-1)=-\frac{5}{3}$.\\
However, with $\Omega_D / \Omega_\Lambda <0$ there is another barrier: the ratio $(H/H_0)^2$ in Eq.\eqref{HSQ} becomes zero for a critical value $z_c$ that solves the quintic equation
\begin{align}
 \frac{\Omega_M}{\Omega_\Lambda} (1+z_c)^5 + (1+z_c)^2 + \frac{\Omega_D}{\Omega_\Lambda}=0 \label{ZCRIT}\,.
 \end{align}
 The critical value $z_c$ is evaluated and discussed in Sect.\ref{RESULTS}.
\end{remark}

\section{Observational data\\ and statistical analysis } \label{sec:data_analysis}
In this section, we detail the datasets used in our analysis as well as the analytical inference of fundamental cosmological quantities using the CKG model.

\subsection{Supernovae type Ia}
SNeIa are a specific type of supernova explosion occurring in binary systems after the accretion of mass of a white dwarf from its companion star. Given their intrinsic mechanisms, it is possible to infer their intrinsic luminosity (and thus their absolute magnitude $M$) from the shape of the detected light curve. For this reason, they are considered standard candles and are of fundamental importance in observational cosmology \cite{Riess 98}. \\
The difference $m-M$ among the 
observed and the absolute magnitudes is related to the luminosity distance $D_L$ (in Mpc) by the relation
$m -M =5\log_{10}D_L+25$. The luminosity distance is defined as
\begin{gather}
D_L (z) =\frac{(1+z)}{H_{0}\sqrt{\mid\Omega_{k}\mid}}S_{k}
\left(\sqrt{\mid\Omega_{k}\mid}\int_{0}^{z}\frac{dz'}{H(z')/H_0}\right)\\
S_{k}(x)=\begin{cases}
\sin x & k=1\\
x & k=0\\
\sinh x & k=-1
\end{cases}
\end{gather}

\subsection{The Pantheon+ dataset}
Refs.\cite{Scolnic 22, Brout 22} present one of the most recent compilations of spectroscopically
confirmed type Ia Supernovae. It contains 1701 light curves of 1550 
SNeIa in the redshift range $0<z<2.3$. It is worth noticing that
77 of them are in galaxies that host Cepheids in the
range $0.00122<z<0.01682$ (see \cite{Malekjani 25}). It comprises
18 different samples, loosely defined as data sets
produced by single SNe surveys in definite periods of time. This sample presents distance moduli 
$\mu=M-m$ anchored to the Classical Cepheids of SH0ES collaboration \cite{Riess 21}, 
as well as magnitudes already corrected for the Tripp \cite{Tripp 98} formula.

\subsection{The Union3  dataset}
This compilation is a comprehensive dataset of 2087 Type Ia Supernovae
from 24 samples in the range $0.01<z<2.26$, reported in
\cite{Rubin  25}. Up to the time of this analysis, the complete dataset has not yet been released. 
Only the 22 bins representative of the entire catalog have been used in our analysis.

\subsection{BAO \& DESI}
In the early Universe before decoupling, the density fluctuations of 
baryonic matter  produced acoustic waves traveling
in the primordial plasma. These are the baryonic acoustic oscillations (BAO). 
The maximum distance covered
from the Big Bang to the time of decoupling (drag epoch)
is the {\em comoving sound horizon} at the drag epoch $r_d$:
\begin{equation}
r_{d}=\int_{z_{d}}^{+\infty}\frac{c_{s}(z')dz'}{H(z')}\label{eq:comoving sound horizon}\,.
\end{equation}
Here, $c_{s}(z)$ is the speed of sound in the photon-baryon
fluid, and $z_{d}\approx1080$ (the last scattering surface). Before the last scattering it is 
\[
c_{s}(z)=\dfrac{c}{\sqrt{3}}\dfrac{1}{\sqrt{1+\dfrac{3\mu_{b}}{4\mu_{R}(1+z)}}}
\]
 where $\mu_{b}$ is the baryonic energy density.  The results of the Planck
collaboration \cite{Planck 2018} give $r_{d}=(147.21 \pm0.48)$\,Mpc. The maximum distance $r_d$ travelled by these acoustic waves left an imprint which manifests in  correlations on the large scale structure of the Universe. It  can be measured to infer the cosmological parameters.  In this case, the BAO can be considered distance rulers. For instance, 
the BAO measurements from DESI DR2 \cite{Abdul DESI release 2} are
described by the following quantities:\\  
- the transverse comoving distance $D_{M}/r_d$, with
\[
D_{M}(z)=c\int_0^z\frac{dz'}{H(z')}
\]
- the Hubble horizon $D_{H}/r_{d}$
\[
D_{H}(z)=\dfrac{c}{H(z)}
\]
-  the angle averaged distance $D_{V}/r_{d}$
\[
D_{V}(z)=\left(zD_{M}(z)^{2}D_{H}(z)\right)^{1/3}
\]
These measures target tracers of bright galaxy samples, luminous
red galaxies, quasars, and the Lyman-$\alpha$ forest.

\subsection{SDSS DR16 }
The 16$^{th}$ Data release from the SDSS contains the extended Baryon Oscillation Spectroscopic Survey (eBOSS) data \cite{Alam 2021}, which includes 
 all data from eBOSS and its predecessor, the Baryonic Oscillation
Spectroscopic Survey. The BAO scale is measured both in the auto\-correlation
of Lyman-$\alpha$ absorption and in its cross-correlation with
341,468 quasars with redshift $z_q >1.77$.

\subsection{CMB}
The CMB data used in our analysis come from the Planck Collaboration \cite{Planck 2018}. Planck was a full-sky mission that in 4 years mapped the CMB temperature and polarization power spectra with a precision never achieved before, looking at the anisotropies in these spectra. The latest 2018 data release not only provides data but also built-in likelihoods to be used for cosmological calculations. They already address possible systematics like foreground effects 
and use a plethora of nuisance parameters for the marginalization of the proper, cosmological ones.

\subsection{Sound horizon and CMB first acoustic peak\\ in CKG cosmology}
The value of comoving sound horizon $r_{d}$ in Eq.(\ref{eq:comoving sound horizon})
strongly depends upon the early time expansion rate. \\
For $z>1080$, in a spatially flat spacetime, the terms $\Omega_{\Lambda}$ and $\frac{\Omega_{D}}{(1+z)^{2}}$
in Eq. (\ref{HSQ}) are negligible, while radiation and matter dominate:  
\begin{equation}
\left(\frac{H(z)}{H_{0}}\right)^{2}\approx \Omega_R(1+z)^{4}+\Omega_M(1+z)^{3}
\end{equation}
The radiation energy density $\Omega_R=\Omega_{\gamma}+\Omega_{\nu}$ is the same of the $\Lambda$CDM model.\\ 
The CMB term $\Omega_\gamma$ is evaluated with the present CMB temperature
$T_{0}=2.7255\:K$ so that $\Omega_{\gamma}=5.45\times 10^{-5}$.\\
$\Omega_\nu$
is the neutrino term 
\begin{equation}
\Omega_{\nu}=\frac{7}{8}N_{\nu}^{eff}(4/11)^{4/3}\,\Omega_{\gamma}=0.2271\,N_{\nu}^{eff}\Omega_{\gamma}\,.
\end{equation}
where the effective number of relativistic species is
fixed to $N_{\nu}^{eff}=3.046$ in the standard cosmological model
(see \cite{Ichikawa,Mangano 02,Mangano 05}). Then $\Omega_{\nu}=0.6917\,\Omega_{\gamma}$
giving 
$$\Omega_R=9.22\cdot10^{-5}$$ 
With this
estimate, for $z>1080$, it is $\Omega_{R}(1+z)^{4}>1.25\cdot10^{8}$ and,
with $\Omega_{M}\approx0.3$, it is $\Omega_{M}(1+z)^{3}>3.77\cdot10^{8}$.
Then the dark sector is completely negligible and
the comoving sound horizon can be evaluated with the formula 2 in
\citep{Abdul DESI release 2}.

The first acoustic peak is found at the angular scale (see \cite{Planck 2018}
Sec. 3.1)
\begin{equation}
\theta_{s}=\dfrac{r_{d}}{D_{M}(z_{d})} \label{thetas}
\end{equation}
The value of $D_{M}(z_{d})$ depends strongly upon the late time expansion
rate, where radiation is negligible and the Hubble parameter evolves 
as Eq.(\ref{eq:Hubble late time}). Now $\Omega_D$ and $\Omega_\Lambda$ are the relevant parameters with $\Omega_M$:
\begin{equation}
D_{M}(z_{d})=\frac{c}{H_{0}}\int_{0}^{z_{d}}\frac{dz}{\sqrt{\Omega_{M}(1+z)^{3}+\Omega_{\Lambda}+\dfrac{\Omega_{D}}{(1+z)^{2}}}}
\label{eq:DM(z)}
\end{equation}
The Planck collaboration results \cite{Planck 2018} give $100\,\theta_{s}=\,1.04097\pm0.00046$.
It is observed at the corresponding multipole 
\begin{equation}
\ell\simeq \frac{\pi}{\theta_s}
\end{equation}
Measurements of the first acoustic peak can be found in \cite{Durrer 03,Harrison 00}

\section{Cosmological results}\label{sec:results} 
In this section, we detail the Bayesian methodology followed in our analysis and the main cosmological results.

\subsection{Methodology}
We used a Bayesian approach to derive the posterior probabilities for the cosmological quantities studied in our analysis: $H_0$, $\Omega_M$, and $\Omega_D$. We fixed $\Omega_{\Lambda}=1-\Omega_M-\Omega_D$ and assumed spatial flatness ($\Omega_k=0$).

\begin{figure*}
\centering
\includegraphics[width=0.45\hsize]{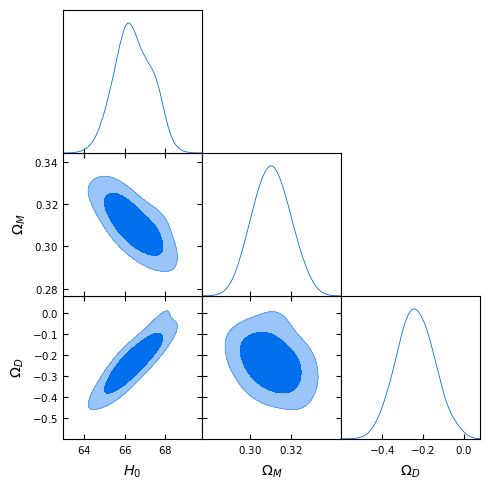}
\includegraphics[width=0.45\hsize]{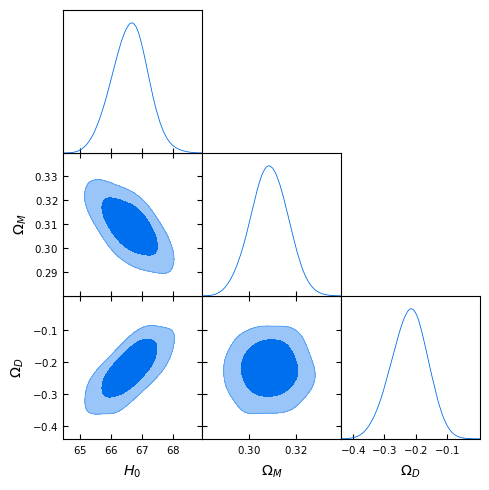}
\includegraphics[width=0.45\hsize]{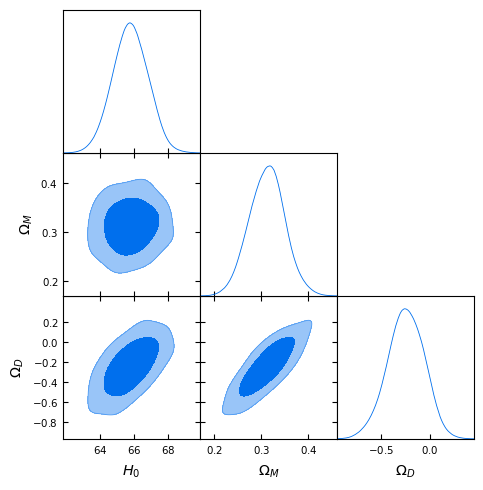}
\includegraphics[width=0.45\hsize]{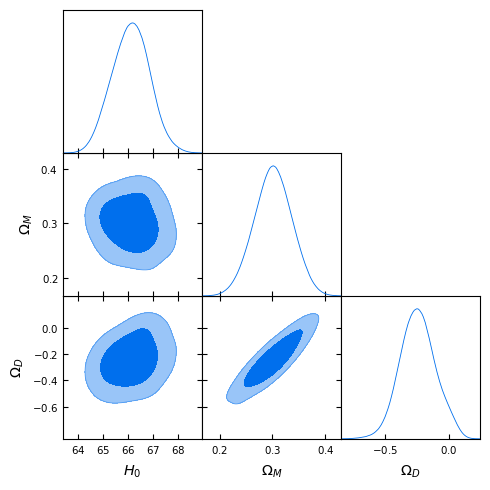}
\caption{Results of the cosmological analysis: DESI+Union3 (top left panel); DESI+Pantheon+  (top right panel); DR16+Union3 (bottom left panel); DR16+Pantheon+ (bottom right panel). The contours represent the $68 \%$ and $95 \%$ probability levels, respectively.} 
\label{Fig_results}
\end{figure*}

Regarding the SNeIa, we used a marginalized chi-squared function to break the degeneracy between $H_0$ and the absolute magnitude $M$. We followed the method outlined in  \cite{Goliath2001, Conley11}. This is defined as 
\begin{equation}
\chi^2_{\text{SN}} = {\sf\Delta}^T \mathbf{C}^{-1} {\sf\Delta} - \frac{[{\sf 1}^T \mathbf{C}^{-1} {\sf \Delta}]^2}{{\sf 1}^T \mathbf{C}^{-1} {\sf 1} }, \label{CHIBAO}
\end{equation}
where $\mathbf{C}^{-1}$ is the inverse of the covariance matrix for the SNeIa datasets, and 
\begin{equation}
{\sf\Delta} = m_{B}^{\text{corr}} - \mu^{\text{th}}
\end{equation}
Here, $m_B^{\text{corr}}$ is the standardized observed magnitude, and $\mu^{\text{th}}$ is the theoretical distance modulus computed from the cosmological model:
$$ \mu^{\text{th}} = 5 \log_{10} \left( \frac{D_L}{\mathrm{Mpc}} \right) + 25. $$
The difference $\sf{\Delta}$ contains an unknown additive offset $\mathcal{M} = M_B - 5\log_{10}H_0 + 25$ which is analytically marginalized  in the likelihood.

Regarding the BAO, we fixed $r_d=147.5$ Mpc, being consistent with the Planck results,  and used the $D_M$ and $D_H$ values of the DR16 dataset. In analogy with Eq.\eqref{CHIBAO} for SNeIa, we introduce a $\chi^2_{\sf BAO}$.\\
The chi-squared functions for SNeIa and BAO are summed and used for the Bayesian computation. For this purpose, we used COBAYA \cite{Torrado21}, a code for Bayesian analysis tailored for cosmological computations, which contains both the datasets used in our analysis as well as the Markov Chain Monte Carlo (MCMC) sampler for the cosmological posterior contours.

For the analysis which includes the CMB, we used two of the official Planck 2018 likelihoods in COBAYA: the low Temperature-temperature (TT) for low coefficients of the power spectra (i.e. the ones associated with the largest angular scales), and the Plik TTTEEE-lite for the rest. We used the lite likelihood because it automatically marginalizes over the nuisance parameters, allowing us to derive only the cosmological ones for our tests. For the Theory part of the code, i.e. the one properly computing the spectra starting from the likelihood and the cosmological parameters, we used CLASS \cite{Lesgourgues11}, a Boltzmann solver able to derive the anisotropies. For the CMB computations, we fixed all the other fundamental cosmological quantities to marginalize only $\Omega_M$, $H_0$, and $\Omega_D$ as in the computations  using SNeIa and BAO. The CMB was used in conjunction with the two other probes to test the reliability of the results, as well as a case where the flatness condition is not set a priori, as we will show.

\begin{figure*}
\centering
\includegraphics[width=0.45\hsize]{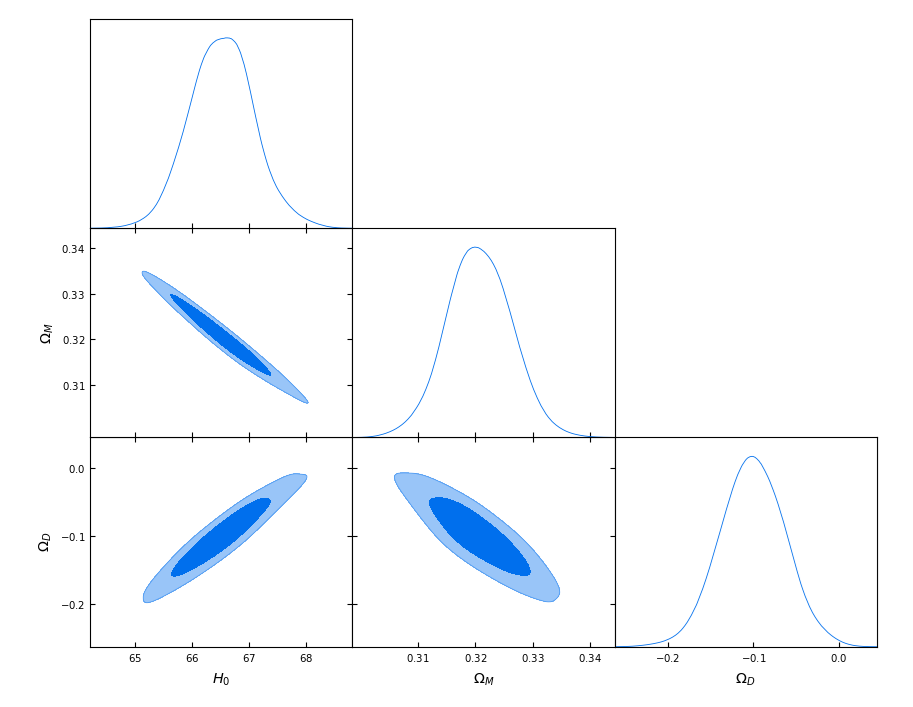}
\includegraphics[width=0.5\hsize]{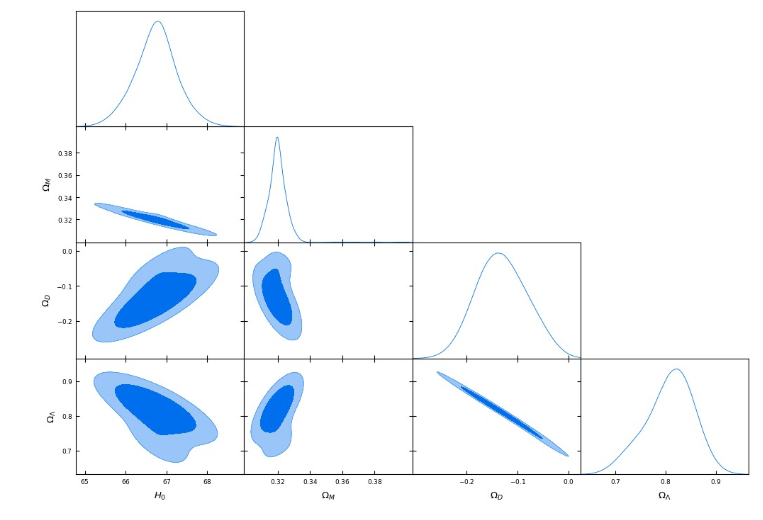}
\caption{Results of the cosmological analysis combining DESI+PantheonPlus+CMB: keeping $\Omega_{\Lambda}$ fixed (left panel), and letting it vary (right panel). The contours represent the $68 \%$ and $95 \%$ probability levels, respectively.} 
\label{Fig_results_CMB}
\end{figure*}

\subsection{Results}\label{RESULTS}
We combined the two datasets of BAO and SNeIa in four different ways and derived the posterior for the cosmological parameters. The main results are shown in Fig.~\ref{Fig_results}, while the mean and one-dimensional standard deviations are shown in Table~\ref{Tab_Results}. \\
The results obtained using also the CMB are shown in Fig.~\ref{Fig_results_CMB} and in 
Table ~\ref{Tab_Results_CMB}. For these, the SNeIA and BAO sets used are the Pantheon+ and DESI, respectively. 
\begin{table}[h]
\centering 
\scriptsize
\begin{tabular}{|c|c|c|c|}
\hline
Datasets &  $H_0$ &$\Omega_M$ & $\Omega_D$  \\ \hline
DESI+Union3 &$66.44 \pm 0.93$ &  $0.310 \pm 0.009$ & $-0.235 \pm 0.095$ \\ \hline
DESI+Pantheon+ &$66.61 \pm 0.58$ &  $0.309 \pm 0.008$ & $-0.221 \pm 0.057$ \\ \hline
DR16+Union3 &$65.77 \pm 1.04$ &  $0.311 \pm 0.037$ & $-0.25 \pm 0.19$ \\ \hline
DR16+Pantheon+ &$66.11 \pm 0.74$ &  $0.301 \pm 0.036$ & $-0.243 \pm 0.14$ \\ \hline
\end{tabular}
\caption{One-dimensional mean and standard deviation for the cosmological parameters derived for the CKG cosmology using 4 different combinations of the SNeIa and BAO datasets.}
\label{Tab_Results}
\end{table}

The results obtained for all the cosmological parameters are each other consistent within $1 \sigma$ in all the considered combinations of datasets.\\
The values for $\Omega_M$ and $H_0$ that were obtained for the CKG cosmological model, are consistent with the results achieved by the Planck collaboration \cite{Planck 2018}: this is an important consistency test confirming the reliability of the results. 

By comparing the contours in Fig.\ref{Fig_results},  it is worth noticing  that their shapes depend on the combination of datasets. $\Omega_D$ and $\Omega_M$ correlate strongly for the sets DR16+Union3 and DR16+Pantheon+, but they do not show the same correlation for  results involving DESI. Although this can be due to the fact that more data are needed in both cases to achieve stable shapes of contours, this is still an interesting difference worth mentioning.

The results for $\Omega_D$ are consistent with what found in \cite{Mantica 24} with the CC+BAO set, presumably because BAO datasets have also been used in the present  work. Specifically, we note that 
$\Omega_D$ is always negative, at least for more than $2\sigma$ levels according to our results. Given that we fixed 
$\Omega_{\Lambda}=1-\Omega_M-\Omega_D$, the ratio $\Omega_D/\Omega_{\Lambda}$ is negative for all our results, thus the discussion in Sec. \ref{sec:Friedmann-equations} applies.

\begin{table}[h]
\centering 
\scriptsize
\begin{tabular}{|c|c|c|c|c|}
\hline
Case &  $H_0$ &$\Omega_M$ & $\Omega_D$ & $\Omega_{\Lambda}$  \\ \hline
Flat & $66.53 \pm 0.57$ &  $0.321 \pm 0.007$ & $-0.101 \pm 0.038$ & . \\ \hline
Not-Flat &$66.74 \pm 0.53$ &  $0.321 \pm 0.009$ & $-0.131 \pm 0.050$ & $0.808 \pm 0.049$ \\ \hline
\end{tabular}
\caption{One-dimensional mean and standard deviation for the cosmological parameters derived for the CKG cosmology using CMB and  the SNeIa and BAO datasets Pantheon+ and DESI.}
\label{Tab_Results_CMB}
\end{table}

We now describe the results obtained using also the CMB in conjunction with SNeIA and BAO. As previously mentioned, the computations have been performed using the CLASS theory code. From its point of view, adding the CKG contribution with $\Omega_D$ is equivalent to add a second cosmological fluid with $w=-5/3$, that acts together with the standard one of the $\Lambda$CDM model, for which we recall $w=-1$. A $w=-5/3$ corresponds to a phantom dark energy ($w<-1$), which in our case complements the standard Equation of State evolution. It is interesting to note that, for high values of the redshift, the phantom contribution becomes negligible with respect to the usual term, while its relative contribution grows with time according to its dependence upon the scale factor.

On the left panel of Fig. \ref{Fig_results_CMB}, we show the results obtained by varying $H_0$, $\Omega_M$, and $\Omega_D$ as in the previous computations. We note how the values for $\Omega_M$ and $H_0$ are consistent with those obtained using only SNeIa and BAO. We also note that $\Omega_D$ is centered around $-0.1$, thus still being negative but closer, albeit significantly different, to 0. We also note how adding the CMB allows us to deduce more clearly the various correlations between parameters. In particular, we note the clear anti-correlation between $\Omega_M$ and $\Omega_D$, and the correlation between $H_0$ and $\Omega_D$.

In the right panel of Fig. \ref{Fig_results_CMB}, we drop the flatness assumption we used up to now and derive $\Omega_{\Lambda}$ alongside the other cosmological parameters. This has been done by imposing $\Omega_K=1-\Omega_M-\Omega_D-\Omega_{\Lambda}$. We note how the results are remarkably consistent with those obtained by imposing flatness, which is important because it means that, if we derive $\Omega_K$ from our computations, we recover the flatness even if we do not impose it. Indeed, by computing $\Omega_K$ from the mean values of 
$\Omega_M$, $\Omega_D$, and $\Omega_{\Lambda}$ shown in Tab. \ref{Tab_Results_CMB}, we get 
$\Omega_K=0.002$, remarkably consistent with a flat universe in the error range. This is an important reliability check for the observational results of CKG cosmology.

In solving Eq.\eqref{ZCRIT} for $z_c$ with the results obtained with the DESI+Union3 datasets, we find $z_c=-0.510$. 
If we apply the same conditions to the dark energy and pressure, eq.\eqref{eq:dark pressure and energy}, we find
\begin{align*}
\mu_D>0 &\qquad if \quad z > -1+ \sqrt{|\Omega_D/\Omega_\Lambda |}\approx -0.496 ,\\
p_D>0 &\qquad if \quad z  > -1+ \sqrt{\frac{5}{3}  |\Omega_D/\Omega_\Lambda |}\approx -0.349.
\end{align*}

Table \ref{Tab_Results_2} reports the calculated values of the deceleration
parameter $q_{0}$ \eqref{QZ} with $z=0$, the parameters $w_{D0}$ and $w_{Da}$ in the
linear approximation  \eqref{JBPinCKG} for the EoS parameter of the dark energy,
the first acoustic peak $\theta_{s}$ of CMB in \eqref{thetas}, and
the ages of the Universe evaluated with the central values of Table
\ref{Tab_Results}. For $\theta_{s}$ we used the value
of comoving sound horizon $r_{d}=147.50$ Mpc as
fixed by DESI DR2 in \citep{Abdul DESI release 2}, and the value $z_{d}=1080$ 
in computing the integral (\ref{eq:DM(z)}).

The results show that the deceleration parameter $q_{0}$ is greater
than the one of $\Lambda$CDM model, and the present value of the
dark energy EoS parameter $w_{0}$ belongs to the quintessence regime.
The values $w_{Da}$ for the linear approximation may be compared
for example with the corresponding ones of the CPL parametrization
in \citep{Abdul DESI release 2} (Sect VII A) or in \citep{Giare24a}.
The calculated  ages of the Universe are fully compatible with Planck
results.

The values of the first acoustic peak in Table~\ref{Tab_Results_2} are very close
 to the best value provided by \citep{Planck 2018}. \\
In our opinion, 
this feature is very important because our analysis, despite considering for this comparison only the  
datasets gathered at low and intermediate redshifts (SNeIa and BAO), is in agreement with
 CMB data, in particular with the Planck data release. In other words, we need not shape the dark sector, and CKG  cosmology seems fully predictive in matching early and late Universe dynamics.
\begin{table}[h]
\centering 
\scriptsize
\begin{tabular}{|c|c|c|c|c|c|}
\hline
Datasets         &  $q_0$                 &     $w_{D0}$  & $w_{Da}$ & $100 \,\theta_s$ & Age (Gy) \\ \hline
DESI+Union3 &$-0.300 $ &  $-0.773 $ & $-0.609$& $1.0454 $ & 13.78 \\ \hline
DESI+Pantheon+ &$-0.315 $ &  $-0.787 $ & $-0.563$&$1.0460$ & 13.77\\ \hline
DR16+Union3 &$-0.283 $ &  $-0.758 $ & $-0.657$&$1.0368 $ & 13.89\\ \hline
DR16+Pantheon+ &$-0.305$ &  $-0.768 $ & $-0.625$&$1.0284 $ & 13.95\\ \hline
\end{tabular}
\caption{The deceleration parameter \eqref{QZ} at $z=0$, the parameters $w_{D0}$ and $w_{Da}$ in \eqref{JBPinCKG}, the angle $\theta_s$ in \eqref{thetas}, the age of the Universe, evaluated at the central values of Table \ref{Tab_Results}.}
\label{Tab_Results_2}
\end{table}

\begin{table}[h]
\centering 
\scriptsize
\begin{tabular}{|c|c|c|c|c|}
\hline
Datasets         &  $q_0$                 &     $w_{D0}$  & $w_{Da}$ &  Age (Gy) \\ \hline
Flat &${\bf -0.417 } $ &  ${\bf -0.901 }$ & ${\bf -0.228}$&  {\bf 13.79}\\ \hline
non-flat &${\bf -0.385} $ &  ${\bf -0.871} $ & ${\bf -0.308}$&{\bf 13.70}\\ \hline
\end{tabular}
\caption{ The deceleration parameter \eqref{QZ} at $z=0$, the parameters $w_{D0}$ and $w_{Da}$ in \eqref{JBPinCKG}, the age of the Universe, evaluated at the central values of Table \ref{Tab_Results_CMB}.}
\label{Tab_Results_3}
\end{table}

\section{Growth of perturbations in CKG cosmology} \label{perturbations}
In this section we investigate the equations governing
the growth of perturbations, in the framework of spherical collapse illustrated by Abramo \textit{et al.} \cite{Abramo07}. The procedure has been used in several theories of extended gravity \cite{Farsi22,Farsi23,Silveira94,Usman23,Ziaie20}.\\
In a spatially flat FLRW Universe with a pressure-less dust, the $\Lambda $ and the dark terms, a
local perturbation of the matter density changes the background value $\mu_M$ to $\mu_M (1+\delta_M)$. 
The parameter $\delta_M$ is the ``density contrast'' \cite{Peebles}. 
Its evolution was derived in \cite{Mantica 24}, in the linear approximation: 
 \begin{align}
 0=&\frac{d^2\delta_M}{dz^2}(1+z)^2  \nonumber \\
 & + \frac{1}{2}\frac{d\delta_M}{dz} (1+z)  \frac{\Omega_M(1+z)^5 - 2 \Omega_\Lambda (1+z)^2
 - 4 \Omega_{D} }{\Omega_M (1+z)^5 + \Omega_\Lambda (1+z)^2 +\Omega_{D} } \nonumber\\
& - \frac{\delta_M}{2} \frac{3\Omega_M(1+z)^5 +8 \Omega_D}{\Omega_M (1+z)^5 + \Omega_\Lambda (1+z)^2 +\Omega_D} 
 \label{DELTAALL}
\end{align}
The following analytic solutions were studied:   

1) $\Omega_D =\Omega_\Lambda =0$ (GR evolution):
\begin{align}
\delta_{M,GR} (z) = c_1(1+z)^{3/2}+ c_2 (1+z)^{-1} \label{DELTAGR}
\end{align}
The constants $c_1$ and $c_2$ are determined by initial conditions at a reference redshift $z_i\gg1$ (matter dominated era). The requirement that the fluctuation is small and the derivative is negative and small 
(initial growth of structures) rules out $c_1$ as unphysical and poses the ``adiabatic condition''
\begin{align}
\delta'_M (z_i) = - \frac{\delta_M(z_i)}{1+z_i} \label{eq:adiabatic condition}
\end{align}
that is used as a criterion to fix the constants in more general conditions.

2) $\Omega_\Lambda \neq 0$, $\Omega_D=0$ ($\Lambda$CDM, Peebles \cite{Peebles}, Martel \cite{Martel91}). 
With $\alpha = \Omega_M/\Omega_\Lambda $,  
matter dominance means $\Omega_M(1+z)^3\gg \Omega_\Lambda $ i.e. $\alpha (1+z)^3\gg 1$.
The admissible solution is
\begin{align}
\delta_M (z)& = c_3 \frac{1}{1+z}\;
{}_2F_1\left (1,\frac{1}{3}; \frac{11}{6}; -\frac{1}{\alpha (1+z)^3} \right )
\end{align}
In the dominant matter regime the hypergeometric function is around unity i.e. the 
$\Lambda$CDM density contrast becomes GR for large $z$.
For large $z_i$ it fulfills the adiabatic condition \eqref{eq:adiabatic condition}.

\begin{figure}[t]
\begin{center}
\includegraphics*[width=8cm,clip=]{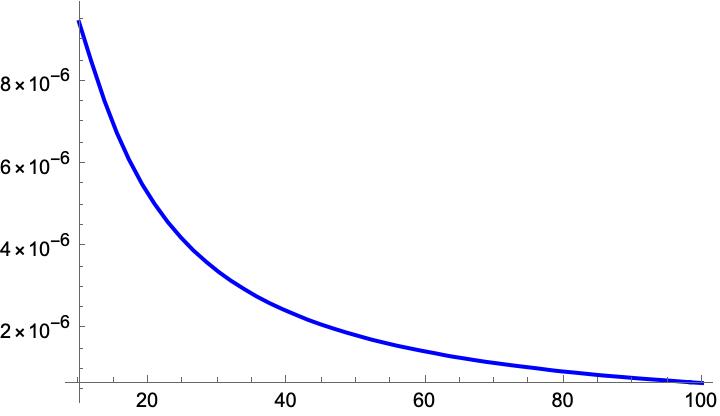}
\caption{\label{delta}  The deviation of the density contrast for CKG from the values in GR, 
$\delta_M(z) - \delta_{M,GR}(z)$. The first one is the numerical solution of \eqref{DELTAALL}, the GR function 
is $c_2/(1 + z)$. 
Both are evaluated for $z>10$,
with initial conditions $\delta_M(400)=0.0001$ and $\delta'_m(400)=-\delta_M(400) /401$.} 
\end{center}
\end{figure}

3) CKG with $\Omega_\Lambda \neq 0$, $\Omega_M+\Omega_\Lambda +\Omega_D=1$.\\

Equation \eqref{DELTAALL} is solved numerically with {\sf NDSolve} of Mathematica13. 
The solution does not show significative difference with the expanding mode of GR in the matter-dominated phase (see Fig.\ref{delta}). 

A robust measurable quantity in red-shift surveys, that is related to the density contrast is the product
$ f(z) \sigma_8 (z)$ \cite{Nesseris 17} (for a simple account of the theory see \cite{Kolb90}). 
$\sigma_8(z)$ is the root mean square mass fluctuation at the scale $R_8=8 h^{-1}$Mpc:
$$  \sigma^2_8(z) =4\pi \int_0^\infty k^2 dk\, \widetilde W^2(k) P_{\delta_M} (k,z)   \label{eq:growth rate}$$
In the integral, $P_{\delta_M} (k,z)$ is the mass power spectrum, i.e. the Fourier transform of the correlator
$\langle \delta_M({\bf x})\delta_M({\bf x'})\rangle $ at redshift $z$. $\widetilde W(k) $ 
is the Fourier transform of a window function $W(r)$ that filters the spatial scale; a choice is $W(r)=3/(4\pi R_8^3)$ if $r<R_8$ and $W(r)=0$ if $r>R_8$.\\
The tension in $\sigma_8 =\sigma_8(0)$ is the discrepancy between the values
measured in the late Universe, that are smaller than the values found
in CMB (early Universe). 
Planck data \cite{Planck 2018} give
$\sigma_{8}=0.8120\,\pm\,0.0073$. 
A complete survey of $\sigma_{8}$
tension is contained in \cite{Periv 22}, Sec. 3.1, together with
a wide list of measures of $\sigma_{8}$ (Table 2 and fig. 21). 

\begin{figure}
\begin{center}
\includegraphics[width=8cm,clip=]{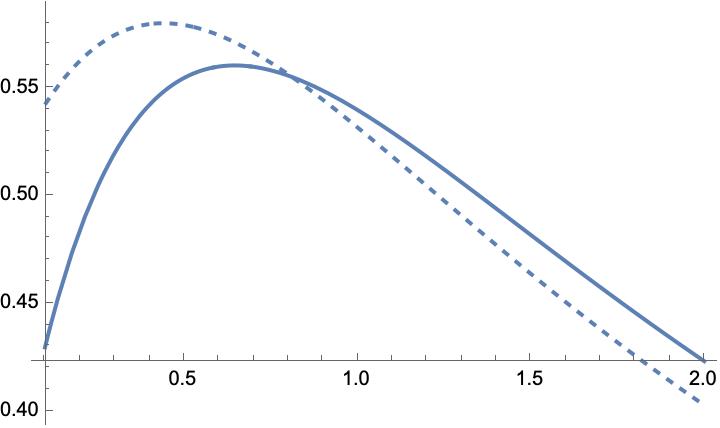}
\caption{\label{GG}  The function $g(z)$, Eq.\eqref{eq:fsigma8 z}, for $\Lambda$CDM (dashed) and CKT (full).}
\end{center}
\end{figure}

$f$ is the growth rate of linear perturbations:  
\begin{align*}
f = \frac{d\log \delta_M}{d\log a} = -\frac{1+z}{\delta_M(z)} \frac{d\delta_M}{dz}, 
\end{align*}
In the linear perturbation regime, $\delta_M (z)$ and $\sigma_8 (z)$ are proportional  (see Eq.2.21 of
\cite{Nesseris 08}). Then, we consider the function
\begin{equation}
g(z) = \frac{f(z)\sigma_{8}(z)}{\sigma_8(0)} = - \dfrac{1+z}{\delta_M (0)}\dfrac{d\delta_M (z)}{dz}\label{eq:fsigma8 z}
\end{equation}
Experimental measures of $f(z)\sigma_8 (z) \propto g(z)$ show a weak maximum, as in Fig.1 of \cite{Benisty21}.\\
The function is plotted for $\Lambda$CDM and CKG in Fig.\ref{GG}. Both curves show a maximum. 
The maximum occurs in a redshift range that is 
borderline of the dominant matter era, wherein the linear evolution equations for $\delta_M(z)$ were obtained. Since the factors $\delta_M(0)$ for $\Lambda$CDM and CKG are very close (respectively $0.0316$ and $0.0295$ with
 the data of Fig.\ref{delta}), the two curves can be compared in the same plot Fig.\ref{GG}. 

\begin{figure}
\begin{center}
\includegraphics[width=8cm,clip=]{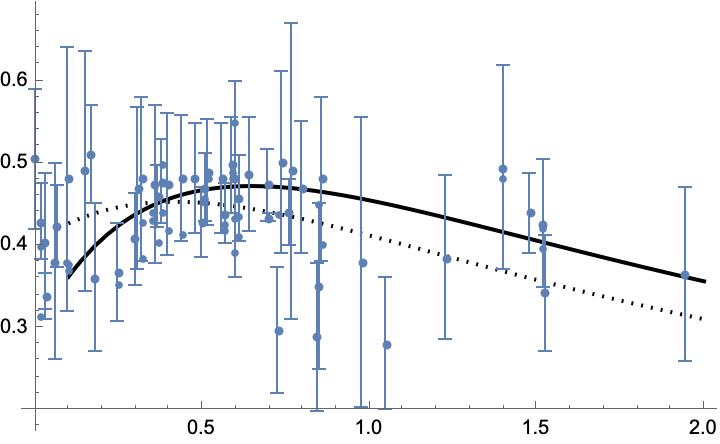}
\caption{\label{GG2}  The function $f(z)\sigma_8(z)$ for $\Lambda$CDM (dotted) and CKG (full), with data points. The error bars are from Table 1 by Benisty, ref.\cite{Benisty21}.}
\end{center}
\end{figure}

%
The function $g(z)$ and datasets for $f(z)\sigma_8(z)$ allow in principle a determination of $\sigma_8$ in CKG.\\
According to the linear theory, Eq.\eqref{eq:fsigma8 z}, the ratio $f(z)\sigma_8(z)/g(z)$ is constant and equal to $\sigma_8$. 
We join two datasets: Benisty (Table 1 with 47 points in \cite{Benisty21})  and Kazantzidis \& Perilovaropoulos (Table II with 63 points in \cite{Kazantzidis 18}), and evaluate $g(z)$ at the same points with \eqref{eq:fsigma8 z}. Different points yield different ratios; the best fitting values $\sigma_8$  (omitting error bar analysis of input data) 
are: $\sigma_8=0.775$ ($\Lambda$CDM) and $\sigma_8=0.843$. With these data, the plots for $g(z)$ are rescaled
to obtain plots for $f(z)\sigma_8(z)$, that are shown with experimental data in Fig.\ref{GG2}. 

\section{Discussion and conclusions}\label{sec:discussion and conclusions}
In this paper we present posterior results for the cosmology derived from CKG with a MCMC methodology and different combinations of cosmological datasets, namely BAO, SNeIa, and the CMB.\\ 
We first considered a flat FLRW spacetime, with parameters $H_0$, $\Omega_R$, $\Omega_M$,  and the novel dark parameter $\Omega_D$ coming from the CKG cosmology. 
The results are consistent with the standard scenario for $H_0$ and $\Omega_M$.\\
We obtain negative values for $\Omega_D$ in all the analysed cases. This is consistent with the previous study on CKG cosmology \cite{Mantica 24}, where a negative value for $\Omega_D$ was found using CC+BAO and a qualitative analysis of $\sigma_8$ was done.  We also complemented our analysis with one case where we drop the a priori flatness assumption, but we recover it from the results, passing another important test for the CKG cosmology.\\
A consequence of $\Omega_D<0$ is the existence of a negative critical value for the redshift $z_c$ that nullifies the ratio $H/H_0$ and thus the second Friedmann equation. \\
The evaluation
of the present time deceleration parameter and the dark energy EoS
parameter confirm a quintessence regime.\\ 
The lookback time is used to evaluate the age of the Universe, and confirms Planck
results.\\ 
Finally, the first acoustic peak of the CMB is estimated, with
remarkable proximity to the best value of the Planck collaboration.

The CKG evolution of the density contrast $\delta_M(z)$ in the linear regime shows
no sensible deviation in the matter dominated-regime from the $\Lambda$CDM or GR solutions.\\
The CKG curve for $f(z)\sigma_8(z)$ shows a maximum that would not properly occur with $\Omega_D>0$, 
in qualitative agreement with datasets.\\
 The present results call for a thorough investigation of the peak of $\sigma_8$ and its deviation from $\Lambda$CDM.\\
In conclusion, it appears that CKG might consistently explain the cosmic expansion up to the
last scattering surface. In a forthcoming paper, we will further detail the model by considering more datasets.
\section*{Data Availability Statement}
All data used in this work are openly available in Refs.\cite{Abdul DESI release 2,Adame 25 DESI,Adame 25 b,Planck 2018,Scolnic 22,Brout 22,Rubin  25,Alam 2021}.
 \section*{Acknowledgements}
S. C.  acknowledges the Istituto Nazionale di Fisica Nucleare (INFN) Sez. di Napoli,  Iniziative Specifiche QGSKY and MoonLight-2  and the Istituto Nazionale di Alta Matematica (INdAM), gruppo GNFM, for the support.
This paper is based upon work from COST Action CA21136 -- Addressing observational tensions in cosmology with systematics and fundamental physics (CosmoVerse), supported by COST (European Cooperation in Science and Technology).
\end{document}